**Electrical and Noise Characteristics of Graphene Field-Effect Transistors: Ambient Effects, Noise Sources and Physical Mechanisms**

S. Rumyantsev[1,2], G. Liu[3], W. Stillman[1], M. Shur[1], and A. A. Balandin[3, *]

[1]Center for Integrated Electronics, Department of Electrical, Computer and Systems Engineering, Rensselaer Polytechnic Institute, Troy, NY 12180 USA

[2]Ioffe Institute, Russian Academy of Sciences, 194021 St. Petersburg, Russia

[3]Nano-Device Laboratory, Department of Electrical Engineering and Materials Science and Engineering Program, Bourns College of Engineering, University of California – Riverside, Riverside, CA 92521 USA

## Abstract

We fabricated a large number of single and bilayer graphene transistors and carried out a systematic experimental study of their low-frequency noise characteristics. A special attention was given to determining the dominant noise sources in these devices and the effect of aging on the current-voltage and noise characteristics. The analysis of the noise spectral density dependence on the area of graphene channel showed that the dominant contributions to the low-frequency electronic noise come from the graphene layer itself rather than from the contacts. Aging of graphene transistors due to exposure to ambient for over a month resulted in substantially increased noise attributed to the decreasing mobility of graphene and increasing contact resistance. The noise spectral density in both single and bilayer graphene transistors either increased with deviation from the charge neutrality point or depended weakly on the gate bias. This observation confirms that the low-frequency noise characteristics of graphene transistors are qualitatively different from those of conventional silicon metal-oxide-semiconductor field-effect transistors.

---

[*] Electronic address: balandin@ee.ucr.edu ; group web—site: http://www.ndl.ee.ucr.edu

## I.     Introduction

Graphene is defined as a planar single sheet of $sp^2$-bonded carbon atoms arranged in honeycomb lattice. Since it first mechanical exfoliation by the Manchester, U.K. – Chernogolovka, Russia research group [1], graphene attracted tremendous attention of the physics and device research communities [2-6]. Specific characteristics of single-layer graphene (SLG), such as its high carrier mobility, up to ~27000 $cm^2V^{-1}s^{-1}$ at room temperature (RT) [1-3], over 200,000 $cm^2V^{-1}s^{-1}$ at cryogenic temperature [4] and very high intrinsic thermal conductivity, exceeding ~3000 W/mK [5-8] near RT, make this material appealing for electronic, sensor, detector and interconnect applications [9-10]. For comparison the, RT electron mobility and thermal conductivity of silicon (Si) are 1500 $cm^2V^{-1}s^{-1}$ and 145 W/mK, respectively. In terms of heat conduction graphene can outperform not only the best bulk materials but also carbon nanotubes (CNTs) [7].

Owing to its extraordinary properties, graphene has been also proposed as material for the future microelectromechanical systems (MEMS) and nanoelectromechanical systems (NEMS) [11]. MEMS/NEMS technology offers miniaturization for realizing significant reduction in weight, size and power consumption. The proposed MEMS/NEMS applications of graphene are facilitated by its planar flat geometry and demonstrated integration with Si and Si/$SiO_2$. Graphene atomic planes suspended across trenches in Si/$SiO_2$ wafers have been used for a variety of experimental studies [5-6]. Graphene and few-layer graphene (FLG) appear to be excellent materials for fabrication of NEMS resonators [11]. Graphene is extremely strong and stiff compared to Si based materials. Its chemical inertness and tunable electronic properties are also advantageous. Graphene membranes of macroscopic size (~100 μm in diameter) have sufficient stiffness to support extremely large loads, millions of times exceeding their own weight [12]. The latter is a major benefit for the MEMS/NEMS applications. The micrometer-size sensors made of graphene are capable of detecting individual molecules owing to its extraordinary high electron mobility and its one-atomic-layer thickness [13].

At the same time, all envisioned electronic, sensor, MEMS/NEMS and system-on-a-chip applications of graphene require low levels of flicker noise (also referred to as $1/f$ noise, excess noise or low-frequency noise), which dominates the noise spectrum at frequencies $f$ below 100 kHz [14]. The flicker noise spectral density is proportional to $1/f^{\gamma}$, where $\gamma$ is a constant close to 1. The unavoidable up-conversion of flicker noise in electronic systems limits many applications. This kind of noise limits the sensitivity of sensors and contributes to the phase noise of microwave oscillators or mixers. Another example is MEMS technology, which has a potential for realizing RF variable capacitors with the performance that is superior to conventional solid-state diodes in terms of non-linearity and losses. In the MEMS thermal shear-stress sensors, the flicker noise and thermal conduction properties are of particular relevance. The sensitivity of graphene sensors may be limited by $1/f^{\gamma}$ noise [13]. Thus the low-frequency noise level in graphene has direct relevance to the development of graphene sensor technology. Investigations of the noise sources in graphene devices might help find ways to 1/f noise reduction.

The effect of the exposure to ambient on the noise characteristics is also important, particularly for the sensor applications of graphene. Even a short-time exposure of metallic or semiconducting CNT devices to ambient conditions results in an order of magnitude larger noise than that in CNT devices in vacuum [14]. As shown below, aging of uncapped graphene FETs under atmospheric conditions produces even stronger change to their low-frequency noise characteristics.

Apart from being a detrimental effect, which has to be reduced, the low-frequency noise can provide valuable information about the electronic phenomena in graphene and potentially can be used for quality control of graphene devices. The low-frequency noise is a useful tool to study transport mechanisms, impurities and defects [15], and to diagnose reliability problems [16]. The low-frequency noise spectroscopy is a valuable tool for reliability testing of interconnects with different geometry [17] and very large scale integrated (VLSI) circuits. The appearance of $1/f^2$ components in the low-frequency $1/f^{\gamma}$ noise spectra from metal interconnects has been linked to electromigration damage [17-18]. Since graphene layers were proposed for interconnect applications [9-10] the knowledge of their low-frequency characteristics may lead to development of a new tool for accessing quality of graphene interconnects.

Previous studies of $1/f$ noise in graphene transistors showed that in terms of the Hooge parameter the level of the $1/f$ noise in graphene is relatively small [19-21]. The noise amplitude in graphene transistors had unusual dependence on the gate bias [19-23]. In this paper we report a systematic study of the noise properties of single (SLG) and bilayer (BLG) graphene field-effect transistors (FETs) before and after their aging under ambient conditions over a month period. We present evidence that the dominant noise contributions in our graphene transistors come from the graphene itself. However, the noise characteristics are strongly affected by the contacts because of the voltage distribution between contacts and the graphene layer.

The rest of the paper is organized as follows. In Section II we briefly describe fabrication of single layer graphene field-effect transistors (SLG-FETs) and bilayer graphene field-effect transistors (BLG-FETs). Section III presents results of the measurements and discussion. Our conclusions are given in Section IV.

## II.       Graphene Device Fabrication and Measurement Procedure

Graphene samples were produced by the mechanical exfoliation from bulk highly oriented pyrolitic graphite (HOPG). All graphene flakes were placed on the standard Si/SiO$_2$ substrates during exfoliation using the standard procedure [1-2] and initially identified with an optical microscope. SLG and BLG sample were selected using micro-Raman spectroscopy through the 2D/G'-band deconvolution [24-27]. The spectra were measured with a Renishaw spectrometer under 488-nm laser excitation in the backscattering configuration [26-28]. We have described in detail the procedure of counting the number of atomic layers in graphene using Raman spectroscopy elsewhere [25-28]. The micro-Raman inspection of the graphene flakes used as the transistor channels was repeated after the fabrication steps to ensure that there was no damage. (The damage to graphene lattice reveals itself through appearance of the disorder D peak in the Raman spectrum [27-28].)

We used Leo1550 electron beam lithography (EBL) to define the source and drain areas through the contact bars with the help of pre-deposited alignment marks. The 10 nm Cr/ 100 nm Au metals were sequentially deposited on graphene by the electron-beam evaporation (EBE). In this design, the degenerately doped Si substrate acted as a back gate. The micro-Raman inspection was repeated after the fabrication and electrical measurements to make sure that graphene's crystal lattice was not damaged.

The current – voltage (I-V) characteristics were measured using a semiconductor parameter analyzer (Agilent 4156B). The characteristics were studied for the pristine (as soon as fabricated) and aged graphene transistors. For the purpose of aging the transistors were kept in ambient environment for over a month time without any voltage applied. The low-frequency noise was measured in a frequency range from 1 Hz to 50 kHz at RT. The graphene transistors were biased in a common source mode at source-drain bias $V_{DS} = 50$ mV. The voltage-referred electrical current fluctuations $S_V$ from the load resistor $R_L$ connected in series with the drain were analyzed by a SR770 FFT spectrum analyzer.

### III.    Noise Measurements Results and Discussion

#### A.   Current-voltage characteristics of graphene transistors

Figure 1 shows two examples of the input current-voltage characteristics of graphene transistors before and after aging. The charge neutrality point was within the range from 10 to 40V for all examined transistors. Aging led to the shift of the charge neutrality point and decrease of the current. Note that as a result of aging the gate-voltage position of the charge neutrality point can either increase or decrease depending on a particular device. The inset shows a typical device structure. The input current-voltage characteristics are often used to estimate the carrier mobility in semiconductor and graphene FETs. There are two commonly used methods for extracting the carrier mobility in the FET channels.

The effective mobility, $\mu_{eff}$, is determined from the channel resistance and given by the following equation

$$\mu_{eff} = \frac{L_g}{R_{eff} \, C_g \, (V_{GS} - V_t) W} , \qquad (1)$$

where $V_{GS}$ is the intrinsic gate-to-source voltage, $V_t$ is the threshold voltage, $L_g$ is the transistor gate length, $C_g$ is the gate capacitance per unit area, $W$ is the gate width, $R_{eff} = \frac{R_{ds} - R_C}{1 - \sigma_0 (R_{ds} - R_C)}$, $\sigma_0$ is the conductivity at the voltage corresponding to the charge neutrality point, $R_C$ is the sum of the drain and source contact resistances, and $R_{ds}$ is the measured drain to source resistance. (All our measurements were performed in linear regime at very small current so that $V_{GS} \approx V_{gs}$, where $V_{gs}$ is the external gate-to source voltage.)

The so-called field-effect mobility, $\mu_{FE}$, is determined from the transconductance, $g_{m0}$, in the linear regime and is given as

$$\mu_{FE} = \frac{g_{m0}}{C_g \, (V_{ds} - I R_C)} \frac{L_g}{W} , \qquad (2)$$

where $V_{ds}$ is the drain-source voltage. In the linear regime at small drain voltages the internal transconductance can be found as

$$g_{m0} \approx g_m \left( 1 + \frac{R_C}{R_{eff}} + R_C \sigma_0 \right) , \qquad (3)$$

where $g_m$ is the external transconductance.

Ideally, both methods should give the same or close values for the mobility. However, Eq. (2) does not take into account the derivative $d\mu/dV_g$, while $\mu_{eff}$ in Eq (1) depends on the value of the threshold voltage, which is often determined with large uncertainty. For these reasons, the values of $\mu_{FE}$ and $\mu_{eff}$ may differ from each other and from the Hall mobility measured on the same wafer.

A rough estimate for the contact resistance, $R_c$, can be obtained by plotting the drain-to-source resistance, $R_{ds}$, versus $1/(V_g-V_t)$. Extrapolating this dependence to $1/(V_g-V_t)=0$ one gets the sum of the drain and source contact resistance, $R_c$. Figure 2 shows an example of such a dependence for "as fabricated" and "aged" transistors (the threshold voltage is taken equal to the charge-neutrality voltage $V_0$). As one can see, at the high electron/hole concentration $((V_g-V_t)^{-1}$ tends to zero) the contact resistance is an essential part of the drain-to-source resistance. Therefore, it should be included into consideration for the mobility estimates and for the noise analysis. As follows from Figures 1 and 2, aging during a month leads to the increase of the contact resistance by approximately one order of magnitude. For all samples the value of the contact resistance per unit width, $R_{c0}=R_c \times W/2$ (W is the graphen flake width), varied from 0.2 to 2 $\Omega$ mm and from 1 to 10 $\Omega$ mm for the pristine and aged samples, respectively.

Figure 3 shows an example of mobilities determined for a typical graphene FET (device 1) with the current-voltage characteristics shown in Figure 1. The results of the calculations near the charge-neutrality voltage are intentionally not shown because this region is characterized by the highest degree of inaccuracy. For the virgin sample both methods for the mobility evaluation yielded very close values of the hole mobilities. For all other cases difference between $\mu_{eff}$, and $\mu_{FE}$ mobilities did not exceed 30% ($\mu_{eff}$ is not shown for the aged sample for clarity). As seen, aging due to environmental exposure over a month leads to mobility degradation. However contacts degradation during the aging was more severe (see Figure 2). As follows from Figures 2 and 3 both contact resistance and mobilities are not equal for electron and holes. That might be an indication of non-ideality of the graphene layer.

### B. Physical mechanisms of low-frequency noise in graphene transistors

Figure 4 shows an example of the noise spectra measured at different gate-bias voltages for one of the graphene transistors. For all examined devices the noise spectra were close to the $1/f^\gamma$ with $\gamma$=1.0-1.1 depending on the gate voltage and a specific device. When the current was changed by the drain voltage the noise spectral density of the short circuit current fluctuations, $S_I$, was always proportional to the square of the drain current, i.e $S_I \sim I_d^2$. In this sense, the measured spectra were similar to the low-frequency noise in devices made from other materials [29]. In the examined set of devices we did not observe clearly distinguishable bulges in the noise spectra, which would indicate fluctuation processes with well defined time constants. In the transistors made of conventional semiconductors such bulges, which appeared only in some devices, were attributed to the generation − recombination (G-R) noise [30-33]. We have previously found G-R bulges in a few graphene FETs on Si/SiO$_2$ substrates [20].

Figure 5 (a) shows the gate voltage dependence of noise spectral density, $S_I/I_d^2$, for all studied devices. One can see a very large dispersion in the data for the noise spectra density of the examined devices. Due to the difference in the distance between the drain and source contacts and the flake width, the active area of the channel varies within a wide range from 1.5 to 80 $\mu$m$^2$. If noise originates from the graphene channel (and not from the contacts) the noise spectral density, $S_I/I_d^2$, should be inversely proportional to the area of the graphene channel, $W \times L_g$. Figure 5 (b) shows the same data as in Figure 5 (a) but normalized to the gate area. As one can see, the dispersion in noise data from device to device is much smaller for $S_I/I_d^2 \times W \times L_g$ than for $S_I/I_d^2$. This indicates that graphene is the dominant source of the low-frequency noise. The data points in Fig. 5 indicated with the blue symbols correspond to SLG transistors, the rest to BLG transistors. In contrast to Refs. [19, 22], we did not observe noticeable difference in the noise amplitude for SLG and BLG devices. One should note though that direct comparison is complicated by differences in devices design (e.g. channel width in our devices was larger) and characteristics (e.g. mobility in our devices was higher).

Let us now consider the gate voltage dependence of noise spectral density. The analysis of this dependence yields valuable information about the noise sources and mechanisms because the gate voltage changes carrier concentration and Fermi level position. In conventional semiconductor FETs the low-frequency noise is usually analyzed in the framework of the McWhorter model [34]. In this model, the low-frequency noise is caused by the tunneling of the carriers from the channel to the traps in the oxide. Therefore the trap concentration in the oxide is a natural figure-of-merit for the noise amplitude in MOSFETs. The McWhorter model predicts that the normalized noise spectral density, $S_I/I_d^2$, decreases in the strong inversion regime as $\sim 1/n_s^2$ (where $n_s$ is the channel carriers concentration). Any deviation from this law might indicate the influence of the contacts, non-homogeneous trap distribution in energy or space, or contributions of the mobility fluctuations to the current noise [29, 35 - 37].

In Figure 6 open symbols show an example of the noise gate voltage dependences in graphene transistor (open symbols). As seen, noise $S_I/I_d^2$ decreases with the increase of the negative value of $V_g$ and does not depend of on the gate voltage for $V_g>0$. The figure also shows the corresponding channel resistance fluctuations, $S_{Ch}/R_{Ch}^2$. For other examined graphene FETs we found a whole variety of the gate-voltage dependences of the noise spectral density: the noise level either increased or decreased with deviation from the charge neutrality point. In some graphene devices, the noise spectral density had a maximum in the noise gate-voltage dependence (sea also Ref. [19] for the gate voltage dependence of noise in graphene). We found no correlation in the slope of noise spectral density versus the gate voltage dependence with the number of graphene layers.

In order to perform detail analysis of the noise characteristics of graphene FETs we have to take into account the influence of the contact resistance even if the contacts are noiseless. The noise spectral density of the current fluctuations is given by the expression

$$\frac{S_{Id}}{I_d^2} = \frac{S_{Ch}}{R_{Ch}^2} \frac{(R_{DS} - R_C)^2}{R_{DS}^2} \qquad (4)$$

where $S_{Ch}/R_{Ch}^2$ is the relative spectral noise density of the graphene channel resistance, $R_{Ch}=R_{ds}-R_C$. In Figure 6, the filled symbols show the resistance fluctuation noise, $S_{Ch}/R_{Ch}^2$, calculated for the same transistor using Eq. (4). As one can see, the noise increases with increasing electron (hole) concentration. For all examined graphene transistors, including both SLG and BLG, we found that far enough from the charge neutrality voltage the noise $S_{Ch}/R_{Ch}^2$ either increases with the increase of the electron (hole) concentration or weakly depends on the gate voltage. At the voltages within a few volts from the charge neutrality points some devices demonstrated very high amplitude of noise and/or unstable noise behavior. Therefore we conclude that noise in graphene transistors does not comply with McWhorter model [34]. The latter has important implications for practical applications of graphene FETs and calls for development of a special model describing low-frequency noise in graphene devices.

There are two types of the physical models of the *1/f* noise relating this phenomenon to either the carrier number fluctuations or the carrier mobility fluctuations. All these models predict that the low-frequency noise scales inversely proportional to the volume of the device (or area for two-dimensional devices) [29]. However, the dependence of the noise spectral density on the carrier concentration varies in different models. The models, which link the *1/f* noise to fluctuations in the number of charge carriers [34], predict a decrease of the noise spectral density, $S_I/I^2$, with increasing carrier concentration (and the total number of carriers in the sample). The $1/f$ noise caused by the mobility fluctuations can appear as a result of fluctuation of the scattering cross section [29]. Such fluctuations can originate from the change in the scattering cross-section due to tunneling transition between two nonequivalent positions of an atom or a group of atoms, motion of dislocations, capture (release) of electrons or holes by a trap, etc. In this case the noise spectral density for one type of the scattering centers is given by (see [38] and references therein)

$$\frac{S_R}{R^2} = \frac{4N_{t\mu}\tau\,w(1-w)}{1+(\omega\tau)^2}l_o^2(\sigma_2-\sigma_1)^2 \quad , \qquad (5)$$

where $N_{t\mu}$ is the concentration of the scattering centers of a given type, $l_0$ is the mean free path (MFP) of the carriers, and $w$ is the probability for a scattering center to be in the state with the cross-section $\sigma_1$. It follows from Eq. (5) that the relative noise spectral density is proportional to the $l_0^2$ and does not depend either on free carrier concentration or on the total number of carriers in the sample. Integration of Eq. (5) over $\tau$ with the appropriate weight yields *1/f* noise, which is proportional to the $l_0^2$. This equation can be further used to analyze the noise origin in graphene devices.

Even though the low-frequency noise in graphene does not comply with the McWhorter model, we can compare the noise amplitude in graphene FETs and conventional MOSFETs. The straight lines presented in Figure 7 show the McWhorter model predictions for the noise amplitudes calculated for different trap concentrations. The regions between lines 1 and 2 and between lines 2 and 3 correspond to the typical noise levels in regular Si n-channel MOSFETs and in Si MOSFETs with the high-k dielectric, respectively. The hitched horizontal region represents the results for the noise spectral density measured in graphene transistors (compare with Figure 6). It is interesting to note that while at the high carrier concentration the noise in graphene is higher than in typical Si MOSFETs, at low carrier concentration the noise in graphene FETs is on the same order of magnitude or smaller than in Si MOSFETs.

As discussed earlier, aging of the graphene transistors due to the environmental exposure leads to the decrease of the carrier mobility and increase of the contact resistance. The noise measurements of the "aged" transistors have shown a substantial increase in the low-frequency noise. Figure 8 shows the gate-voltage dependence of the noise spectral density of the transistors kept in ambient for about 1month. Since both the contact resistance and mobility degrade as a result of the environment exposure we assume that both contacts and graphene layer itself contribute the *1/f* noise increase. The latter was confirmed by the fact that normalization either by the graphene area or contact width have not resulted in the significant decrease of the data spread for the noise spectral density obtained from all examined devices.

## IV.     Conclusions

We investigated low-frequency noise in graphene field-effect transistors focusing on the effects of environmental exposure and analysis of the noise sources. It was found that at the high carrier concentrations (~$4 \times 10^{12}$ cm$^{-3}$) the contact resistance is a substantial part of the overall device resistance. Therefore it should be taken into account for the mobility and noise analysis.  The exposure of the uncapped graphene transistors to ambient for a period of about one month resulted in substantial increase of the contact resistance and mobility degradation. Through analysis of the noise spectral density dependence on the graphene channel area and gate bias we established that the dominant contributions to noise come from graphene itself. For all examined graphene transistors, both SLG and BLG, we found that noise $S_{Ch}/R_{Ch}^2$ either increases with deviation from the charge neutrality point or weakly depends on the gate voltage. The observed noise behavior is very different from that in conventional Si MOSFETs.

### *Acknowledgments*

The work at UCR was supported, in part, by DARPA – SRC Focus Center Research Program (FCRP) through its Center on Functional Engineered Nano Architectonics (FENA) and Interconnect Focus Center (IFC), and by AFOSR award A9550-08-1-0100 on the Electron and Phonon Engineered Nano and Heterostructures. The work at RPI was supported by the National Science Foundation under I/UCRC "Connection One" and by IFC. This work was also partially supported by RFBR (Grant No. 08-02-00010).

**FIGURE CAPTIONS**

Figure 1: Output current-voltage characteristics showing the drain current vs. gate voltage for two SLG transistors. The drain voltage is $V_d$=100 mV. The characteristics shift with time due to ambient exposure. The inset is a scanning electron microscopy image of a typical graphene device. The scale bar in the inset is 3 μm.

Figure 2: Dependence of the drain-to-source resistance $R_{ds}$ on $1/(V_g$-$V_t)$ for one of the graphene transistors before and after one-month "aging" at ambient conditions. The cross-point of the fitted lines is a rough estimate of the contact resistance $R_C$.

Figure 3: Effective (lines) and field-effect (symbols) mobilities as functions of the gate voltage calculated for a typical transistor. Note strong mobility degradation owing to extended exposure to ambient conditions.

Figure 4: Noise spectral density of the relative short-circuit current fluctuations at different gate voltages. The drain voltage is $V_d$=50 mV.

Figure 5: Relative noise spectral density $S_I/I_d^2$ (a) and area-normalized noise spectral density $S_I/I_d^2 \times W \times L_g$ (b) for the same transistors as functions of the gate bias. Note a decrease in the data spread in the noise spectral density normalized by the area of the graphene channel. The data points indicated with blue symbols correspond to SLG transistors, the rest are BLG transistors.

Figure 6: Gate voltage dependence of the current noise spectral density $S_I/I_d^2$ and the channel resistance fluctuations $S_{Ch}/R_{Ch}^2$. The frequency of the analysis is $f$=10 Hz.

Figure 7: Noise spectral density multiplied by the gate area as a function of gate voltage. The tilted lines are calculated for conventional silicon transistors in accordance with McWhorter model for three different oxide-trap concentrations: 1 - $N_t$=5×10$^{16}$ (cm$^3$eV)$^{-1}$, 2 - $N_t$=10$^{18}$ (cm$^3$eV)$^{-1}$, 3 - $N_t$=10$^{20}$ (cm$^3$eV)$^{-1}$. The shadowed region represents the noise level for graphene transistors. The frequency of the analysis is $f$=10 Hz.

Figure 8: Relative noise spectral density $S_I/I_d^2$ as a function of the gate bias for "aged" graphene transistors.

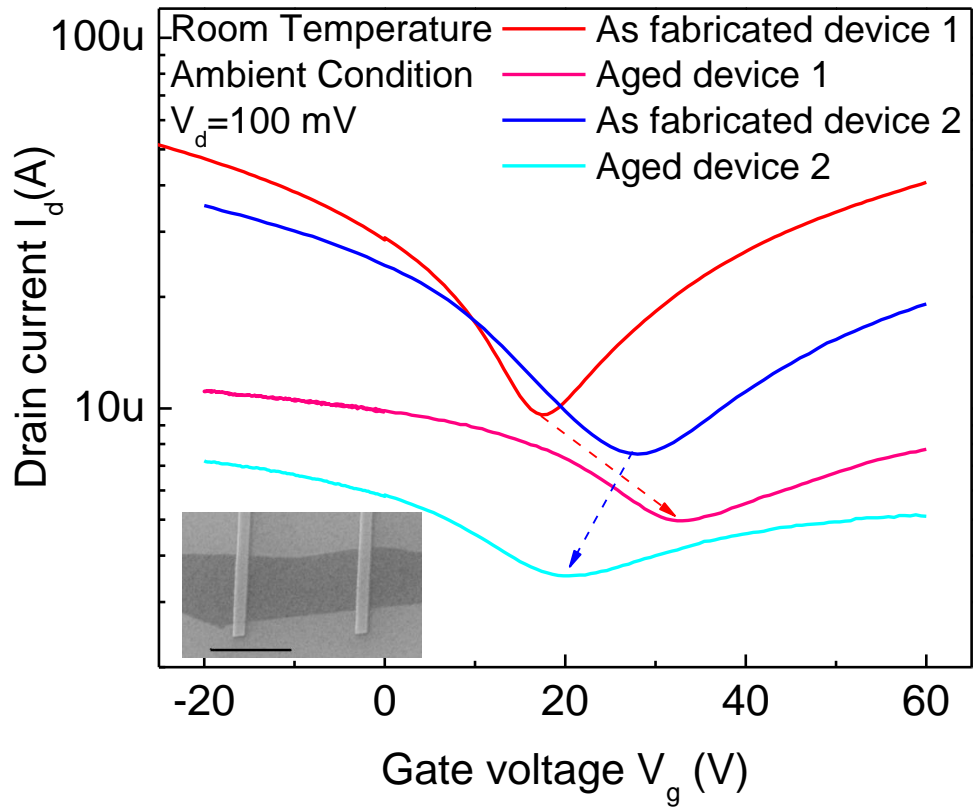

Figure 1

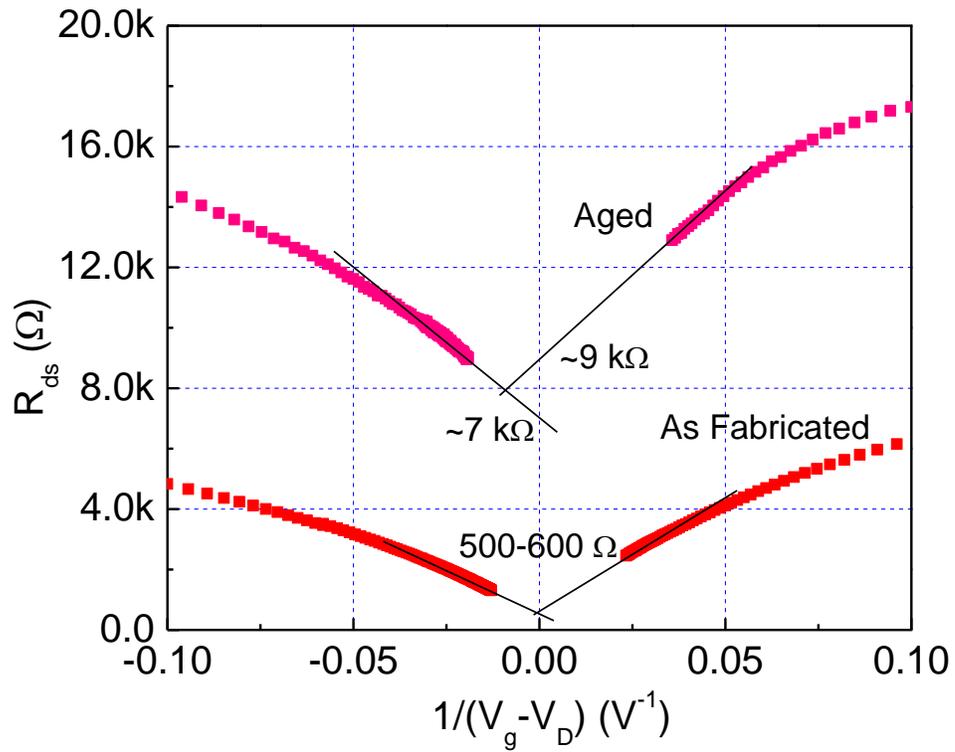

Figure 2

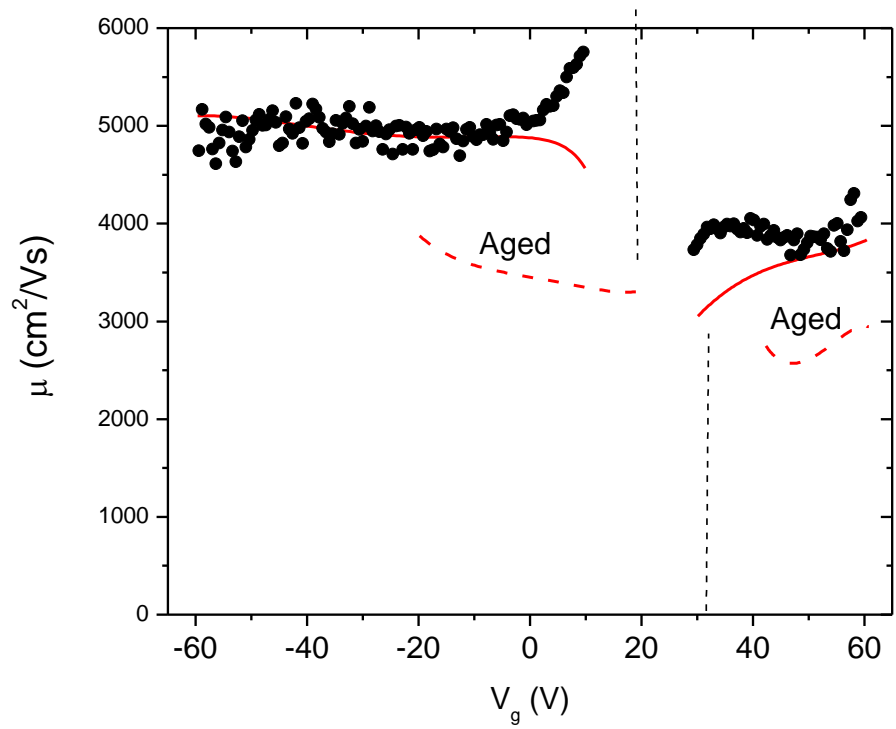

Figure 3

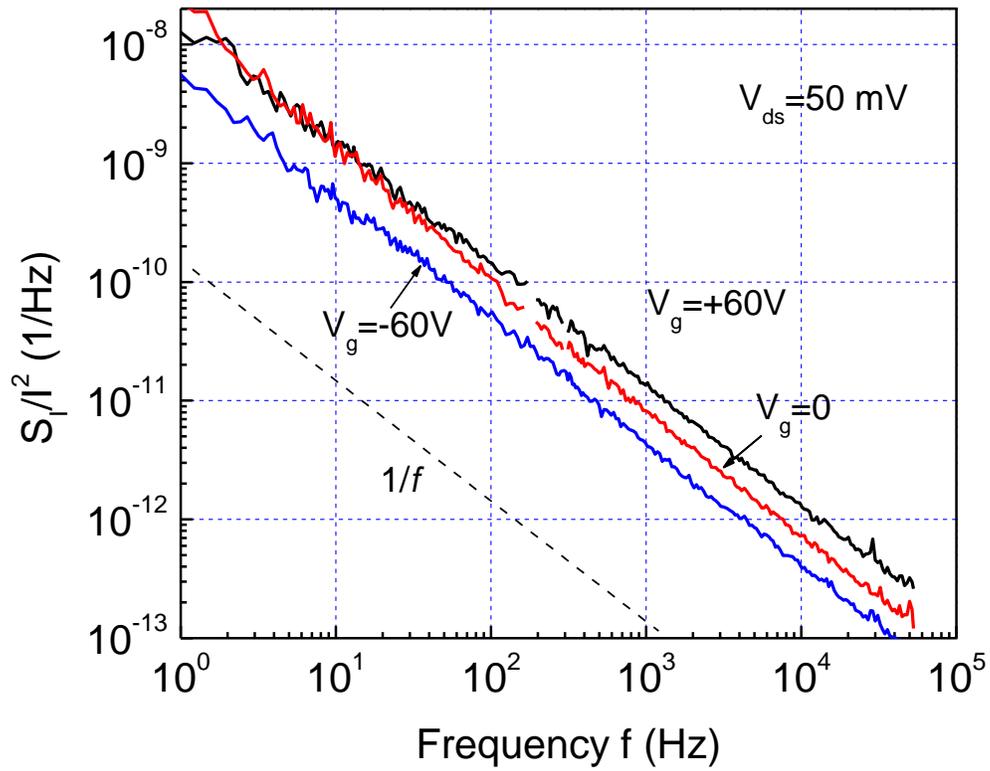

Figure 4

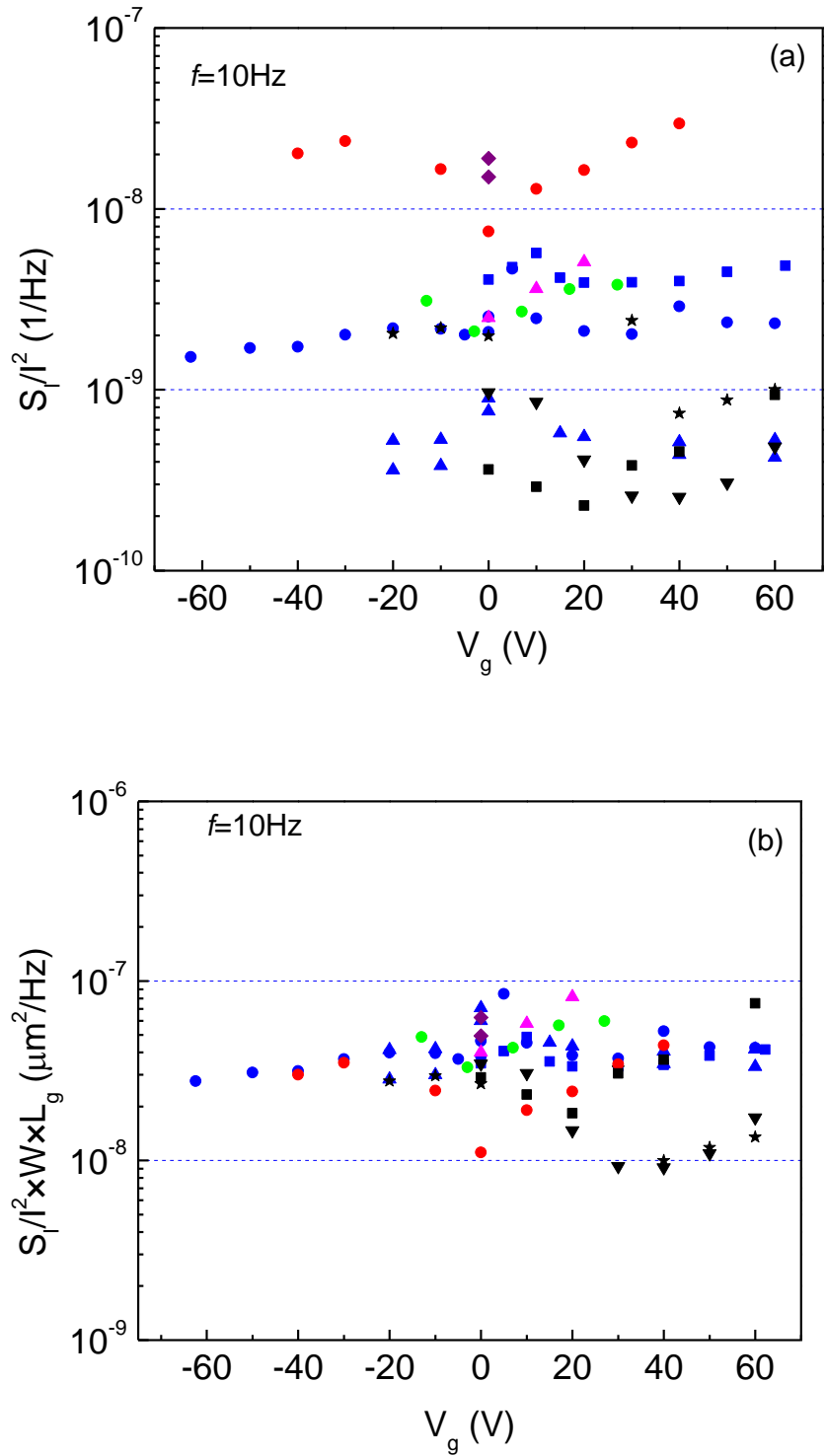

Figure 5

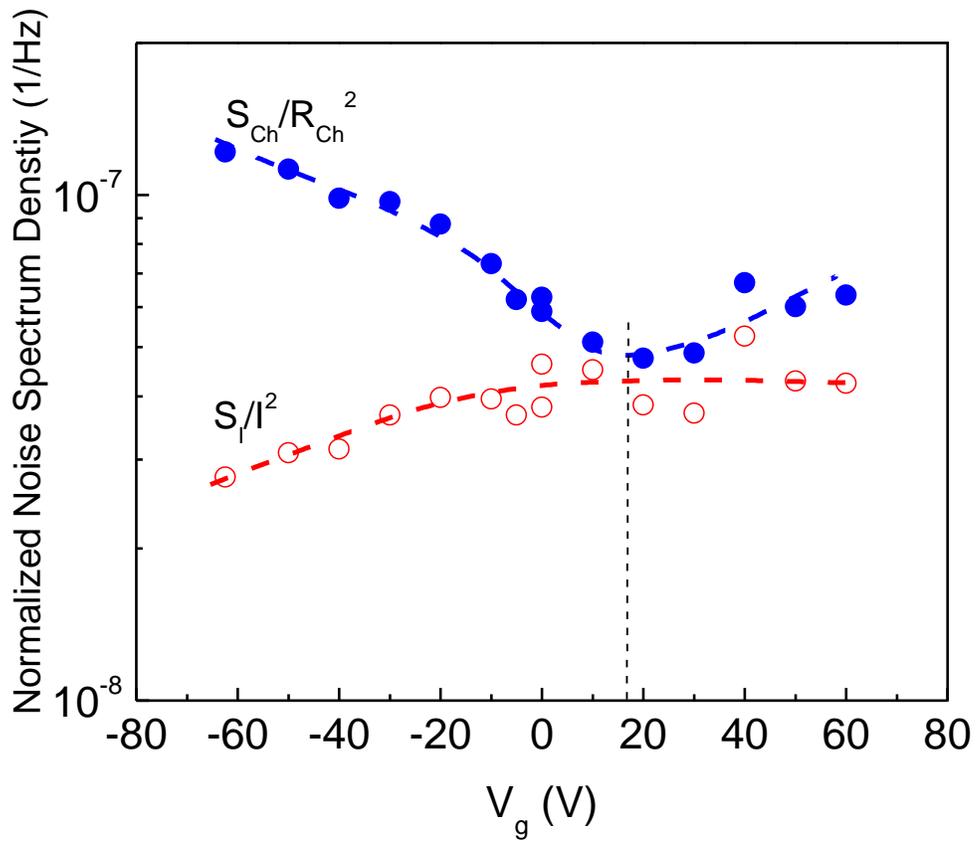

Figure 6

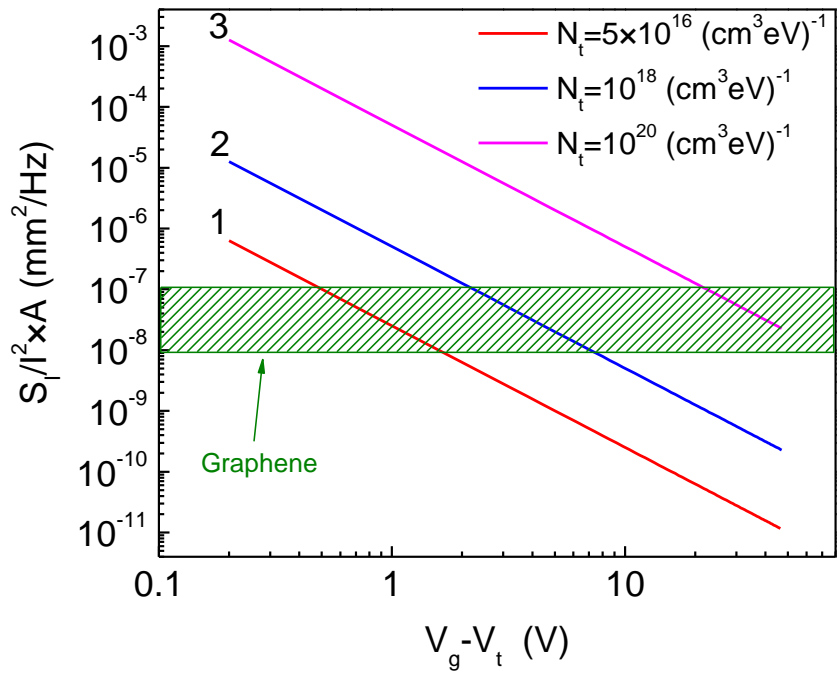

Figure 7

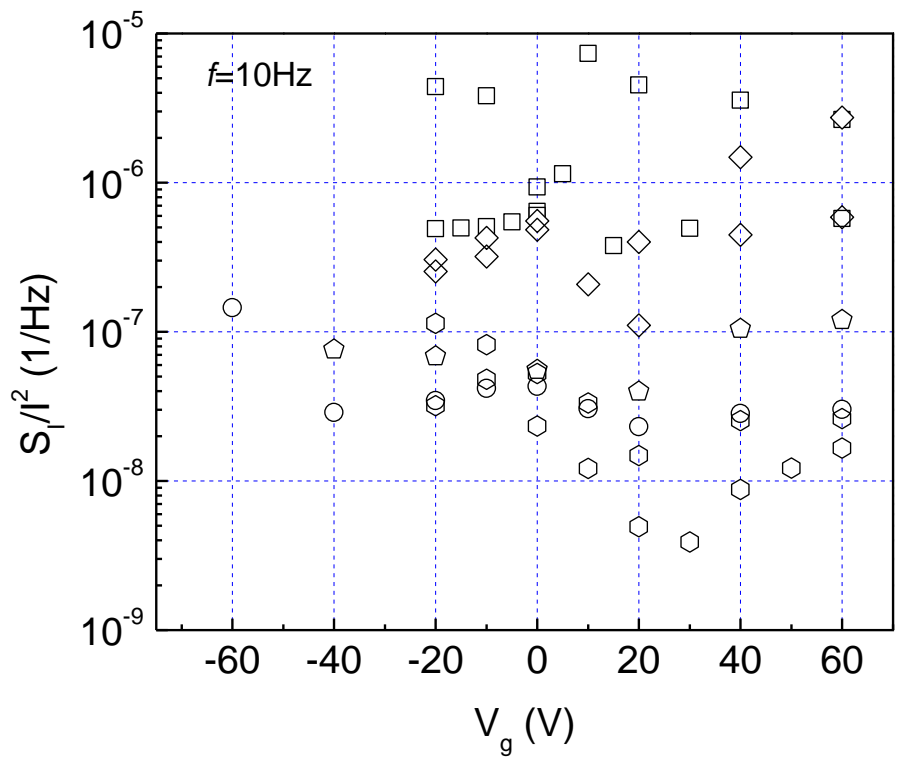

Figure 8